\documentclass[twocolumn,superscriptaddress,amsmath,amssymb,aps,prl,nofootinbib]{revtex4-2}
\usepackage{amsmath}
\usepackage{graphicx}
\usepackage{dcolumn}
\usepackage{color}
\usepackage{physics}
\usepackage{verbatim}
\usepackage{float}
\DeclareGraphicsExtensions{.png .jpg .pdf}

\begin{document}

\title{Enhanced Spin Hall Response From Aligned Kramers-Weyl Points in High Chern Number Semimetals}

\author{C. O. Ascencio}
\affiliation{School of Physics and Astronomy, University of Minnesota, Minneapolis, Minnesota, 55455, USA}
\author{Wei Jiang}\email{jiangw@umn.edu}
%\affiliation{Centre for Quantum Physics, Key Laboratory of Advanced Optoelectronic Quantum Architecture and Measurement (MOE), School of Physics, Beijing Institute of Technology, Beijing, 100081, China }
\affiliation{Department of Electrical and Computer Engineering, University of Minnesota, Minneapolis, Minnesota 55455, USA}
\author{D. J. P. de Sousa}
\affiliation{Department of Electrical and Computer Engineering, University of Minnesota, Minneapolis, Minnesota 55455, USA}
\author{Seungjun Lee}
\affiliation{Department of Electrical and Computer Engineering, University of Minnesota, Minneapolis, Minnesota 55455, USA}
\author{Jian-Ping Wang}
\affiliation{School of Physics and Astronomy, University of Minnesota, Minneapolis, Minnesota, 55455, USA}
\affiliation{Department of Electrical and Computer Engineering, University of Minnesota, Minneapolis, Minnesota 55455, USA}
\author{Tony Low}\email{tlow@umn.edu}
\affiliation{School of Physics and Astronomy, University of Minnesota, Minneapolis, Minnesota, 55455, USA}
\affiliation{Department of Electrical and Computer Engineering, University of Minnesota, Minneapolis, Minnesota 55455, USA}

\date{ \today }
\begin{abstract}
We propose a spin Hall effect (SHE) enhancement mechanism due to Kramers-Weyl point (KWP) alignment in chiral topological semimetals with high Chern numbers (CNs). Through model Hamiltonian calculations, we identify enhancements in the intrinsic spin Hall conductivity (SHC) and the spin Hall angle (SHA). Such enhancements, attributed to a unique high CN KWP energetic alignment and a high degree of SOC-induced band nesting, strongly depend on orbital-orbital interactions. This represents a novel mechanism to enhance SHE, differing from the spin-orbit induced anticrossing mechanism in gapped systems. Guided by this principle, we corroborate our results by means of first-principles calculations and reveal multiple realistic materials with large intrinsic SHCs and even larger SHAs than Pt. 
\end{abstract}

%\pacs{}

\maketitle

\paragraph{Introduction.}The efficient generation of spin currents is a highly sought-after goal within the spintronics community~\cite{Aronov,Datta,Dyakonov,Hirsch,Prins, Uchida,Bauer, Seki,Hirohata, Zhang2021, Amin, Amin2, Duarte2, Baek, Sinova2,Davidson,Liu, MacNeill2017,Shi2,Xue}.
Of particular interest is the intrinsic spin Hall effect (SHE), which relies on the spin-orbit coupling (SOC) of a material, independent of any impurity scattering mechanisms~\cite{Murakami1, Murakami2,Sinova1, Sinova2}. An anomalous velocity due to the Berry curvature associated with a material's band structure is the source of the intrinsic SHE~\cite{Murakami1, Murakami2, Sinova1,Sinova2, Culcer, Gradhand_2012, Berry}. To-date, the mechanism to enhance the intrinsic spin Hall conductivity (SHC) has been primarily attributed to SOC-induced anticrossings~\cite{Zhang2021,Derunova,Pt_SHC} and no general theory to enhance the spin Hall angle (SHA) has been developed.

Analogous to the anomalous Hall effect (AHE) in magnetic systems, where a larger AHE can be achieved in topological systems with higher Chern numbers (CNs)~\cite{QAHE,QAHE2, QAHE3}, high CNs in non-magnetic materials may lead to enhanced SHCs~\cite{McCormick, Shekhar2018, Chang2018}. Chiral topological semimetals (CTSs) are known to host Kramers-Weyl points (KWPs) with the largest possible CN~\cite{MaxChern}, $\mathcal{C} = 4$~\cite{RhSi, Chang2018}, which are suggested to support the longest possible Fermi arc surface states, quantized photogalvanic currents, a unique longitudinal magnetoelectric effect, among other exotic phenomena~\cite{ Tang, chiral2, RhSi, hedgehog1, hedgehog2,Kramers,CoSi_SHC, helicity}.

Previous intrinsic SHE studies on topological materials mainly examined gapped systems with $\mathcal{C} = 1$, where SOC-induced gaps are responsible for the large SHC~\cite{TI_SHC2, Weyl_SHC,Dirac_SHC,Zhang2021,nodal_SHC}. The SHC in CTSs is fundamentally different due to their gapless nature and variable CNs associated with KWPs~\cite{Chang2018,MaxChern,Bradlyn,encyclopedia}. Additionally, due to their relatively low charge conductivity, semimetals with appreciable SHCs may lead to large SHAs. CTSs, therefore, provide an ideal material platform to reveal potentially novel SHE physics in high CN semimetallic systems and to search for highly efficient spin sources for applications in magnetic based memories and in-memory computing technologies~\cite{MacNeill2017,WTe2, Duarte, Sebastian2,Zhang2020}.

In this letter, we propose a distinctive intrinsic SHE enhancement mechanism in high CN CTSs and predict promising material candidates with large SHE. We first uncover positive correlations between the SHC/SHA and CN using multi-orbital chiral semimetal toy models with tunable orbital degeneracy and orbital-orbital coupling (OOC), where the large spin Berry curvature (SBC) near KWPs of the highest CN is the main source of the SHC. Such a SHE enhancement can be ascribed to energetic alignments between KWPs via tunable OOC. We corroborate our findings by means of a space group 198 (SG-198) tight-binding model carrying $\mathcal{C} = 4$, where SOC-induced nesting between linearly dispersing bands that emanate from $\mathcal{C} = 4$ KWPs leads to significant SHE enhancement at alignment. This is further supported by first-principles calculations involving 37 gapless $\mathcal{C} = 4$ materials from the same space group. In particular, two of these materials, BiTePd and BiPdSe, exhibit two of the largest SHAs, surpassing that of Pt, reproducing the salient characteristic features of our SG-198 model. Our study provides a clear strategy for the use of high CN materials for energy efficient spintronics. \\

\paragraph{Chiral Semimetal Toy Models.} We begin by considering a time-reversal ($\mathcal{T}$-) symmetric, single-orbital, CTS model with $\mathcal{C} = 1$ KWPs,  which describes crystalline systems from SG-16~\cite{Chang2018}. The nearest-neighbor (NN) tight-binding  Hamiltonian can be expressed as
\begin{eqnarray}
& \mathcal{H} = \sum_{i=1}^{3} [t_s^{i} \cos \left(k_i\right) \sigma_0 +  t_{SO}^i \sin \left(k_i\right) \sigma_i],
\label{eq1}
\end{eqnarray}
where the first three terms are NN hoppings, $t_s^i$, $i=1$-$3$ denoting the hopping amplitude along $x$, $y$, and $z$ directions. The following three terms are SOC components that define the chirality with SOC strengths $t_{SO} ^i$~\cite{Chang2018}. The Pauli matrices $\sigma_i$ operate on the spin space. SOC lifts the spin-degeneracy throughout the Brillouin zone (BZ) except at time-reversal invariant momenta (TRIM), as enforced by Kramer's theorem, which results in KWPs with $\mathcal{C} = 1$ [Fig.~1(a)]~\cite{SM,Chang2018}.

\begin{figure}
     \centering
         \includegraphics[width=\linewidth]{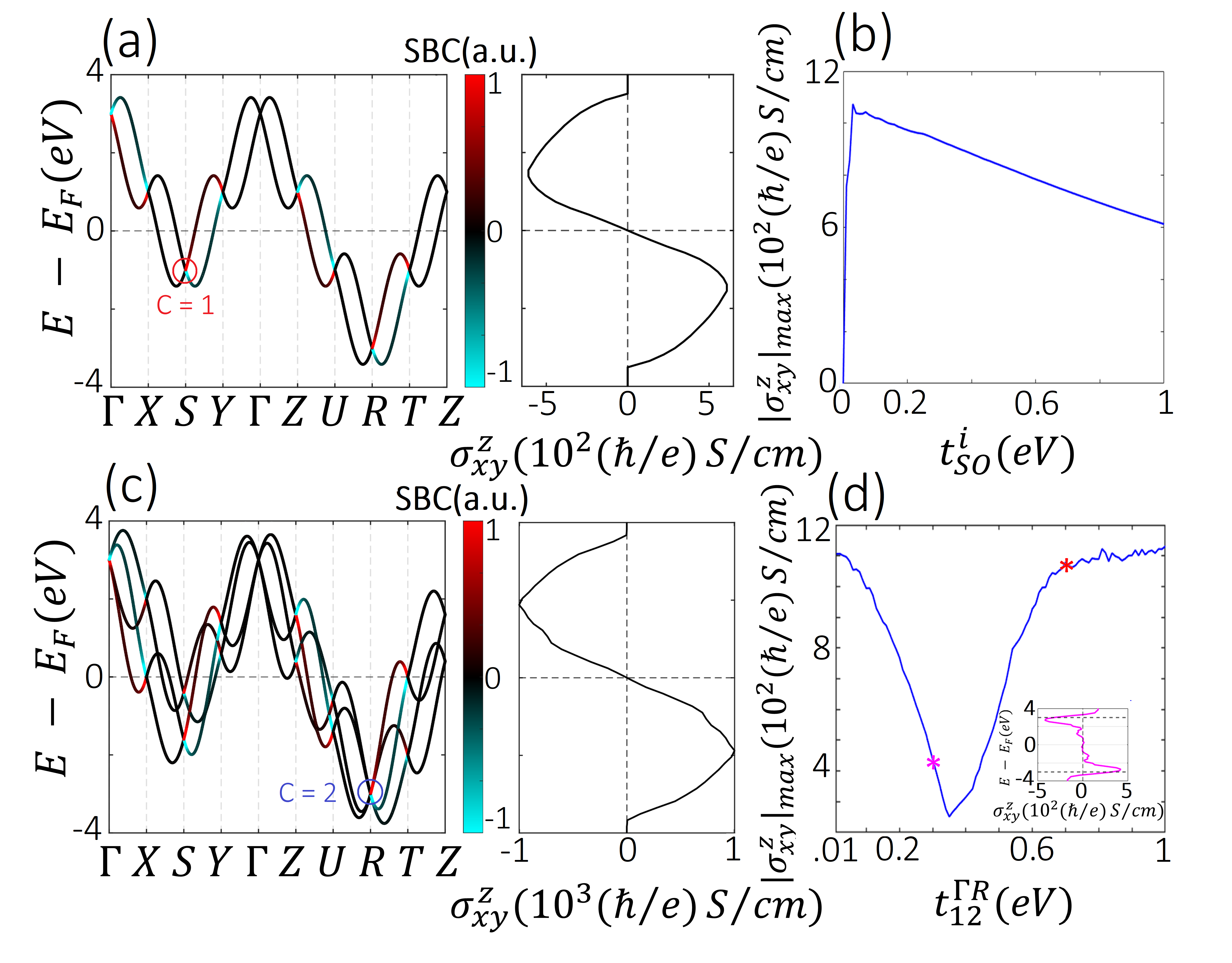}
     \caption{The energy spectrum and SHC for CTS models. (a) $\mathcal{C} = 1$ model band structure with SBC distribution and SHC. Red circle indicates a $\mathcal{C} = 1$ KWP. (b) The effect of SOC on the maximum SHC~\cite{SM}. (c) Same as (a) for $\mathcal{C} = 2$ model ($t_{12}^{\Gamma R} = 0.1$eV). Blue circle indicates a $\mathcal{C} = 2$ KWP. (d) The effect of OOC on the maximum SHC. The inset shows that the SHC peaks at the energies near the $\mathcal{C} = 2$ KWP positions (dashed black lines) for $t_{12} ^{\Gamma R} = 0.3$eV (magenta asterisk). Red asterisk corresponds to alignment of high CN KWPs with a subset of $C = 1$ KWPs at $t_{12} ^{\Gamma R} = 0.67$eV~\cite{SM}}
\end{figure}

We calculate the energy-dependent SHC based on the Kubo formula that integrates the SBC of the occupied bands throughout the BZ~\cite{Sinova1, Sinova2, Guo, Pt_SHC,Gradhand_2012}. The maximum SHC with the selected hoppings is about $|\sigma _{xy} ^z |\sim 610 (\hbar /e)$S/cm [Fig.~1(a)]~\cite{SM}. The SBC distribution shown in Fig. 1(a) reveals that the main contribution to the SHC comes from bands near the KWPs~\cite{SM}. We also studied the influence of SOC on the SHC and plotted the maximum SHC as a function of SOC strength [Fig. 1(b)]. We observe a peak SHC at low SOC followed by a monotonic decrease at larger SOCs due to further SOC-induced band splitting~\cite{Jiang2021,SM}.

To understand the correlation between the SHC and CN in semimetallic systems, we build upon Eq. (1) to construct a four-band model that supports KWPs with $\mathcal{C} = 2$. We double the orbital degeneracy within the unit cell, which is common in chiral crystals, and incorporate  spin-conserving, $\mathcal{T}$-symmetric, SG-16 symmetric OOCs~\cite{SM},
\begin{equation}
\centering
 \mathcal{H} =
\begin{bmatrix}
H_1 & H_c\\ 
H_c^{\dagger} & H_2 
\end{bmatrix},
\label{eq2}
\end{equation}
with two distinct diagonal blocks $H_{n = 1,2} =\sum_{i=1}^{3} [t_{s,n}^{i} \cos \left(k_i\right) \sigma_0 + t_{SO,n}^i \sin \left(k_i\right) \sigma_i]$ and one off-diagonal term, $H_c$, which defines OOC. For simplicity, and consistency with the previous model, the OOC matrix, $H_c = t_{12}^{\Gamma R}[3 \cos \left(k_z\right)  + 2 \cos \left(k_y\right) - 5 \cos \left(k_x\right)] I$, is chosen to preserve electron-hole, time-reversal, and SG-16 crystalline symmetries~\cite{SM}, where $I$ is the second-order unit matrix and $t_{12}^{\Gamma R}$ represents the OOC strength. The form of $H_c$ guarantees $\mathcal{C} = 2$ KWPs at $\Gamma$ and $R$~\cite{SM}. 

Before turning on OOC, this model represents a trivial duplication of the two-band model in Eq. (1), which consists of four-fold degenerate points (i.e., two overlapping $\mathcal{C} = 1$ KWPs) at all TRIMs~\cite{SM}. OOC energetically separates most of these degeneracies into two $\mathcal{C} = 1$ KWPs. Importantly, the four-fold degeneracies at $\Gamma$ and $R$ are preserved due to the coexistence of orbital degeneracy, $\mathcal{T}$-symmetry, and the form of the OOC. The calculated energy-dependent SHC [Fig. 1(c)]~\cite{SM} exhibits a similar trend to the $\mathcal{C} = 1$ model. The maximum SHC for  SOC strengths comparable to the $\mathcal{C} = 1$ case is nearly twice that of the $\mathcal{C} = 1$ results [Fig. 1(c)], as expected since the SHC is proportional to the band degeneracy.

To investigate the effect of OOC on the SHC, we plot the maximum possible SHC as a function of OOC strength, $t_{12}^{\Gamma R}$ [Fig. 1(d)]. Increasing $t_{12}^{\Gamma R}$ initially leads to a decrease in the maximum SHC due to an increase in the energetic misalignment between $C = 1$ KWPs that smears out their SBC contributions~\cite{SM}. On the other hand, the contribution from energetically fixed KWPs at $\Gamma$ and $R$, with $\mathcal{C} = 2$, remains almost unchanged with variable $t_{12}^{\Gamma R}$ [see Fig. 1(d) inset]~\cite{SM}. Interestingly, further increase in $t_{12}^{\Gamma R}$ leads to an increase in the SHC [Fig. 1(d)] due to an energetic alignment between KWPs from different TRIMs~\cite{SM}. A similar analysis with $\mathcal{C} = 2$ KWPs at $\Gamma$ and $S$, which breaks the electron-hole symmetry, exhibits similar behavior~\cite{SM}.

We further investigate systems with higher CNs, i.e., $\mathcal{C} = 3$ and $4$, using a six and eight-band model Hamiltonian by increasing the orbital degeneracy to three and four orbitals, respectively~\cite{SM}. Similar OOC terms are incorporated while preserving the symmetries of SG-16 to generate higher CN systems within the same framework~\cite{SM}. Further analysis confirms a positive correlation between CN and SHC for a large parameter space of these models~\cite{SM}.

The correlation between SHA and CN under variable OOC is less obvious, since the effect of band degeneracy is factored out. To investigate this, we calculate the difference of SHCs and SHAs at $E_F = 3$eV (energy of high CN KWP at $\Gamma$) between $C = 2$ and $C = 3$ models for $0.01$eV $\leq t_{12} ^{\Gamma R} \leq 0.5$eV, as displayed in Fig. 2. Interestingly, we find that concurrent positive correlations between CN, SHC, and SHA can be achieved for a specific range of OOC strengths, as highlighted in Fig. 2(a)~\cite{SM}. To unravel the underlying source of these correlations, we plot the band structures of the $C = 3$ model~\cite{SM} at the SHC difference minimum and maximum [Fig. 2(a) top panel]. The band structure for the maximum difference [Fig. 2(b)] exhibits good alignment between the high CN KWP at $\Gamma$ and  two other KWPs (SBC sources) at $S$ and $Z$, as indicated by the red dashed lines, while the band structure for the minimum difference [Fig. 2(c)] shows an obvious energetic misalignment between them. Thus, it is an energetic KWP alignment that leads to constructive contributions to both the SHC and SHA. Similar results are observed between other CN systems~\cite{SM}.

\begin{figure}
\centering
\includegraphics[width=\linewidth]{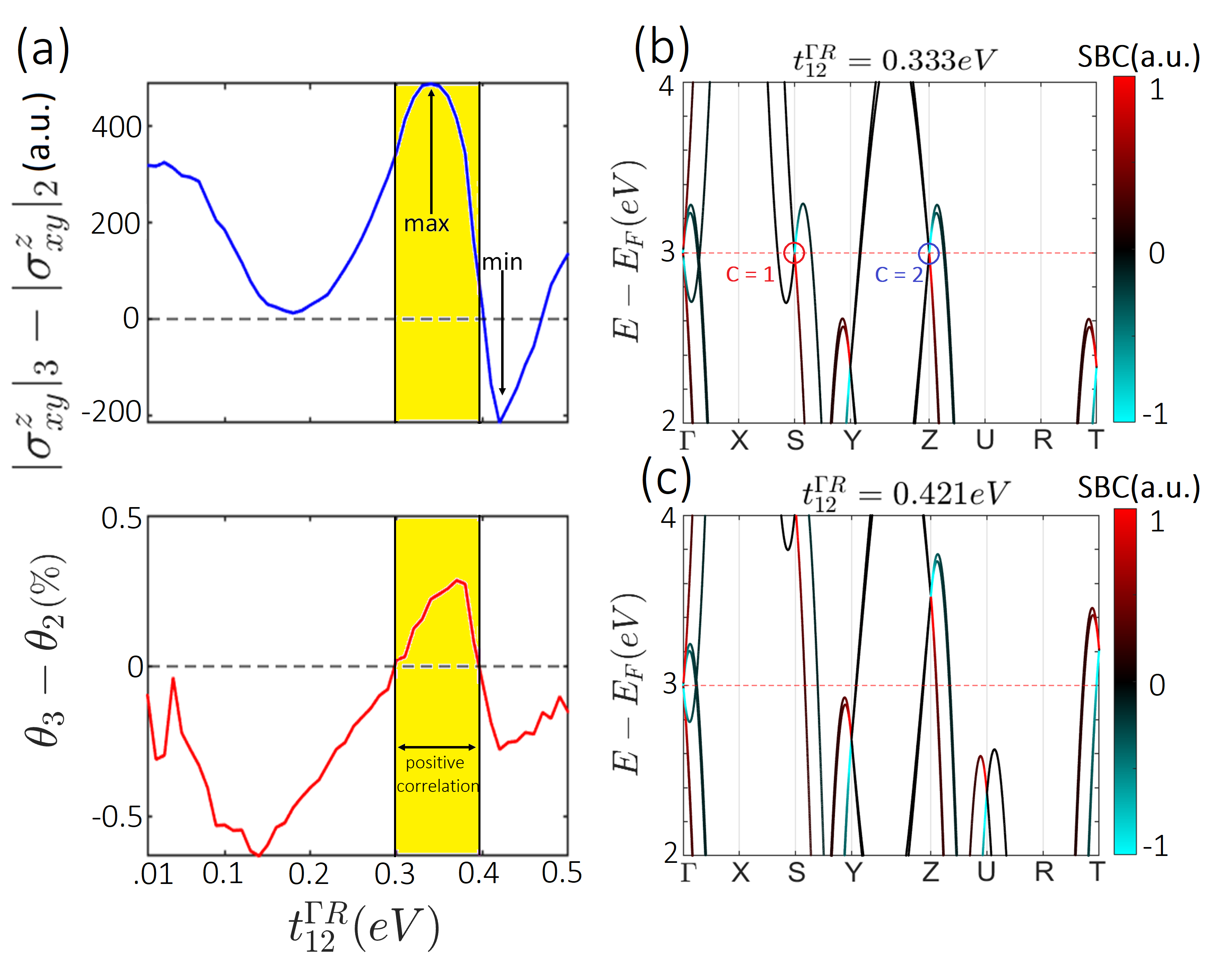}
\caption{KWP energetic alignment and CN-SHE correlation. (a) Difference in SHCs (upper panel) and SHAs (lower panel) between $C = 2$ and $C = 3 $ models for 0.01eV $\leq t_{12}^{\Gamma R} \leq $0.5eV. (b) and (c), $C = 3$ model band structure with KWP alignment [maximum in (a)] at $t_{12} ^{\Gamma R} = 0.333$eV and misalignment [minimum in (a)] at $t_{12} ^{\Gamma R} = 0.421$eV, respectively. Range of OOC strengths that support concurrent positive correlations between CN, SHC, and SHA is highlighted.}
\label{figure2}
\end{figure}

These results indicate that gapless higher CN systems can concurrently have higher SHCs and SHAs under KWP alignment, which is tunable through OOC. We note that a trivial increase in the orbital degeneracy in the absence of OOC would render a nearly constant SHA with variable CN~\cite{SM}. Our models, therefore, hint at the importance of OOC in determining the SHC and SHA of these distinct CN systems, and thus provides a feasible approach for optimizing the SHA in gapless high CN materials. 

We note that the Hamiltonians used thus far rely critically on the specific form of the OOC, which is artificial and may not represent realistic material systems. Fortunately, such simple artificial models can be used to describe realistic chiral systems with an appropriate choice of OOC consistent with the symmetries of a specific chiral space group. We elaborate on this point in the highest CN system and demonstrate an enhanced SHC and, more importantly, SHA enhancement with high CN KWP alignment.\\

\emph{Chiral Semimetal With $\mathcal{C}$ =  4.} Building upon Eq. (1), we construct an eight-band model by quadrupling the orbitals in the system. Rather than artificially placing all the orbitals at the same Wyckoff position, we evenly distribute them by constructing a face-centered cubic (FCC) unit cell [Fig. 3(a)]. The NN interactions, as described by Eq. (1), become next-nearest neighbor (NNN) interactions, and the orbital-orbital terms describe NN interactions. The effective 8 $\times$ 8 Hamiltonian can be written as~\cite{SM}
\begin{equation}
\centering
\begin{aligned}
 \mathcal{H} = \epsilon I + H_{\textnormal{NNN}} + H_{\textnormal{NN}}  + H_{\textnormal{NNsoc}},
 \end{aligned}
\label{eq2}
\end{equation}

where the first term is an on-site potential and the second describes NNN interactions, as inherited from Eq. (1). The following two terms define NN spin-conserving interactions and NN symmetry-preserving SOCs, respectively (See SM for more details). We note that, apart from the on-site potential and NNN SOC terms, this model coincides with the one used by Chang, et. al~\cite{RhSi}.

The band structure, energy-dependent SHC, and SHA at high CN KWP alignment are presented in Fig. 3(b). The spectrum supports a non-trivial four-fold degeneracy at $\Gamma$ and an eight-fold one at $R$, both of which carry $\mathcal{C} = 4$~\cite{SM}. The bands emanating from the $\mathcal{C} = 4$ KWPs contribute sizably to the SHC as displayed by the SBC distribution [Fig. 3(b)], which is a consequence of a combination of the high band degeneracy and SOC-induced band nesting. The energy dependent SHC forms a camel-back type structure [Fig. 3(b) center panel] where two large peaks (red and magenta asterisks) are generated by large, positive SBC contributions from six nested bands which, in part, form the $\mathcal{C} = 4$ KWP at $R$. The negative SBC energetically near the $\mathcal{C} = 4$ KWPs gives rise to the deep valley in the SHC around $E = E_F$ (blue asterisk). The energy-dependent SHA [Fig. 3(b) right panel] consists of three peaks, two of which coincide with the large peaks in the SHC. The central SHA peak is a consequence of the deep valley in the SHC and the low density of states around $E = E_F$~\cite{SM}.

\begin{figure}
     \centering
          \includegraphics[width=\linewidth]{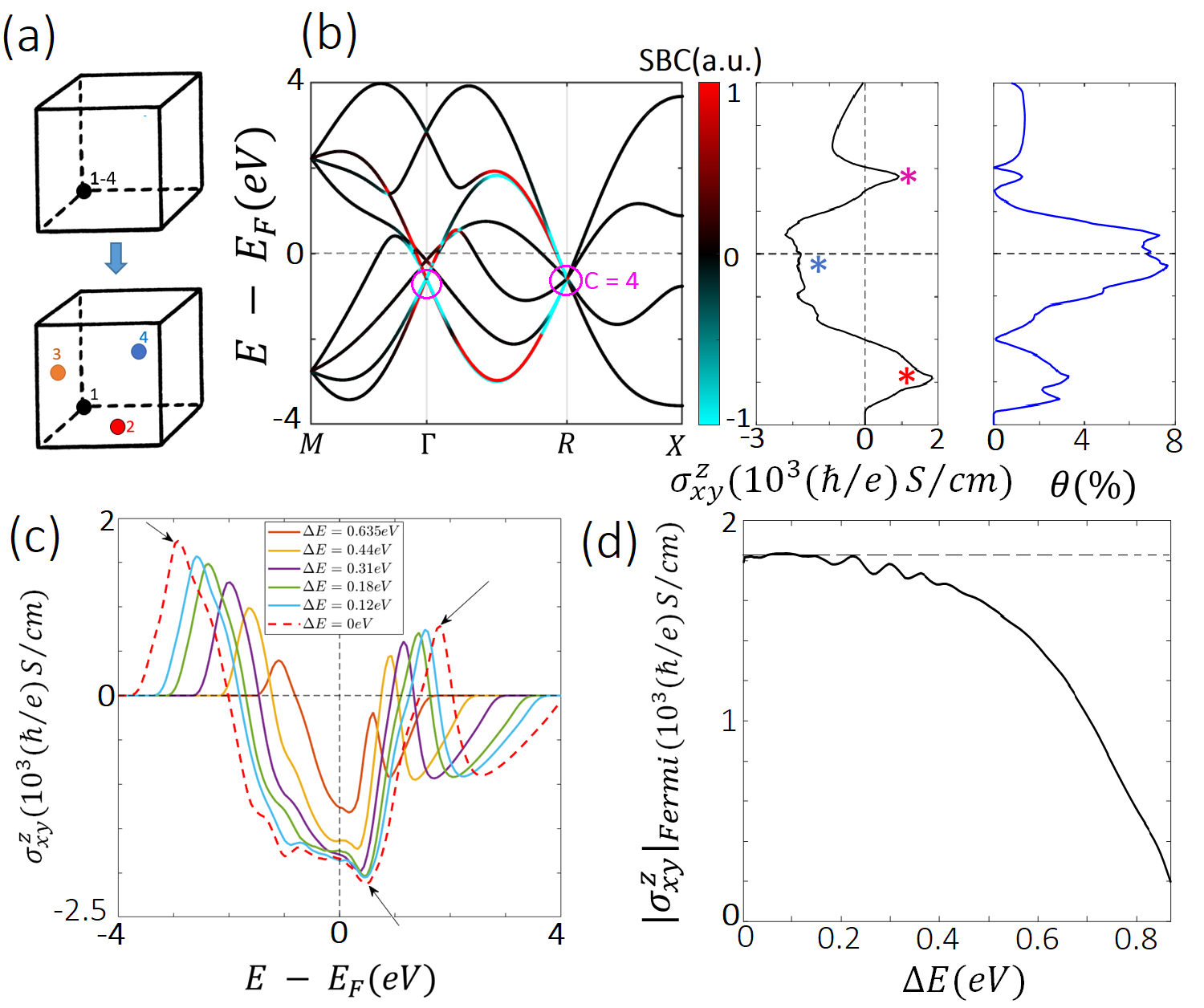}
      \caption{SHE of $\mathcal{C} = 4$ system. (a) Unit cell of SG-16 and SG-198 models. (b) Band structure of SG-198 model at $\mathcal{C} = 4$ KWP alignment with SBC distribution, energy-dependent SHC with camel-back structure, and SHA($\theta$)~\cite{SM}. Magenta circles indicate $\mathcal{C} = 4$ KWPs. Red and magenta asterisks indicate peaks associated with band nesting whereas the blue one indicates a deep valley from large SBC near aligned $\mathcal{C} = 4$ KWPs. (c) Evolution of energy-dependent SHC with changes in $\mathcal{C} = 4$ KWP separation ($\Delta E = E _{\Gamma} - E_R = 0$ corresponds to alignment). Arrows indicate enhancements in peaks and deep valley. (d) Evolution of SHC at $E = E_F$ with changes in $\mathcal{C} = 4$ KWP separation. }
\end{figure}

The influence of $\mathcal{C} = 4$ KWP separation ($\Delta E = E _{\Gamma} - E_R$), which is tuned via OOC~\cite{SM}, on the energy-dependent SHC is also studied[Fig. 3(c)]. At large $\mathcal{C} = 4$ KWP separation (small OOC) the band structure is similar to the trivial case where two $\mathcal{C} = 2$ KWPs overlap at each TRIM~\cite{SM}. Decreasing $\Delta E$ (increasing OOC) leads to an increase in the SHC magnitude, which is attributed to the constructive SBC contributions from bands emanating from $\mathcal{C} = 4$ KWPs. We also examined the evolution of the SHC at $E = E_F$ for a larger range of $\Delta E$ [Fig. 3(d)] and find that the SHC saturates around alignment~\cite{SM}. These results indicate that, with respect to significantly misaligned $\mathcal{C} = 4$ KWPs, SHE enhancement can be achieved by means of improved energetic alignment. %A similar analysis was done for the maximum ISHC over the full bandwidth of the model, which shows that the maximum possible ISHC also occurs at $C = 4$ KWP alignment~\cite{SM}.
%To further corroborate the positive correlation between ISHC and CN, we performed a statistical analysis of the calculated ISHCs for 260 real CTSs~\cite{Zhang2021}, which agrees with our model Hamiltonian results~\cite{SM}.\\

\emph{SHC and SHA of SG-198 CTSs.}
Guided by the preceding principles, we focus on material systems with $\mathcal{C} = 4$, i.e. CTSs from SG-198, to identify materials with large SHCs and SHAs. We studied 37 materials within SG-198 with non-magnetic ground states and the required chiral crystal structure~\cite{Zhang2019, Chang2018,Symmetry_node} within the context of density functional theory~\cite{SM}. We elaborate in detail on two representatives, BiTePd \& BiPdSe (see SM for 35 additional materials~\cite{SM}), which fulfill the $\mathcal{C} = 4$ KWP alignment criterion described~\cite{SM}. Their crystal structure has four atoms per element in the unit cell with Bi and Te(Se) forming opposite rotating chains along the $[111]$ direction[Fig. 4(a)]. Their relativistic band structures support a four-fold KWP at $\Gamma$ and a six-fold KWP at $R$ [Fig. 4(b) and 4(c) left panels], which are protected by the presence of three non-symmorphic two-fold screw axes, a three-fold rotation axis, and $\mathcal{T}$-symmetry~\cite{RhSi}. 

The energy-dependent SHCs for an energy window of $2$eV around the Fermi level are shown in the right panels of Fig. 4(b) and 4(c). The Fermi level SHC for BiTePd (BiPdSe) is about $|\sigma _{xy} ^z | = 893 ~(\hbar /e)$S/cm ($|\sigma _{xy} ^z | = 590 ~(\hbar /e)$S/cm)~\cite{Pt_SHC,SM}. Importantly, we observe a characteristic camel-back structure in the SHC similar to the SG-198 model results [Fig. 3(b) center panel] with the corresponding features indicated by colored asterisks for comparison [Fig. 4(b) and 4(c) right panels]. For both representative materials, the large peak at the Fermi level, which is enhanced by KWP alignment (we further corroborate this mechanism via strain engineering~\cite{SM}), is a consequence of the large SBC contributions from SOC-induced band nesting of six linearly dispersing bands that emanate from the $\mathcal{C} = 4$ KWP at $R$. Additionally, because the energy gradient of these bands is large near the Fermi level, the corresponding density of states can be relatively small in comparison to Pt and may lead to larger SHAs.

Finally, the calculated SHAs ($\theta=\frac{2e}{\hbar}\frac{\sigma _{xy} ^z}{\sigma _{yy}}$) and SHCs ($\sigma _{xy} ^z$) for 37 materials from SG-198 (and fcc Pt:  $ \theta _{Pt} = 0.028$~\cite{SM} ) at the Fermi level  are displayed in Fig. 4(d). Two of the largest possible Fermi level SHAs correspond to our two representative materials (BiTePd: $\theta = 0.063$ and BiPdSe: $\theta = 0.053$). There are additional CTSs from this group with larger SHAs than Pt, which strongly suggests their promise in spintronics applications.\\

\begin{figure}
          \centering
         \includegraphics[width=\linewidth]{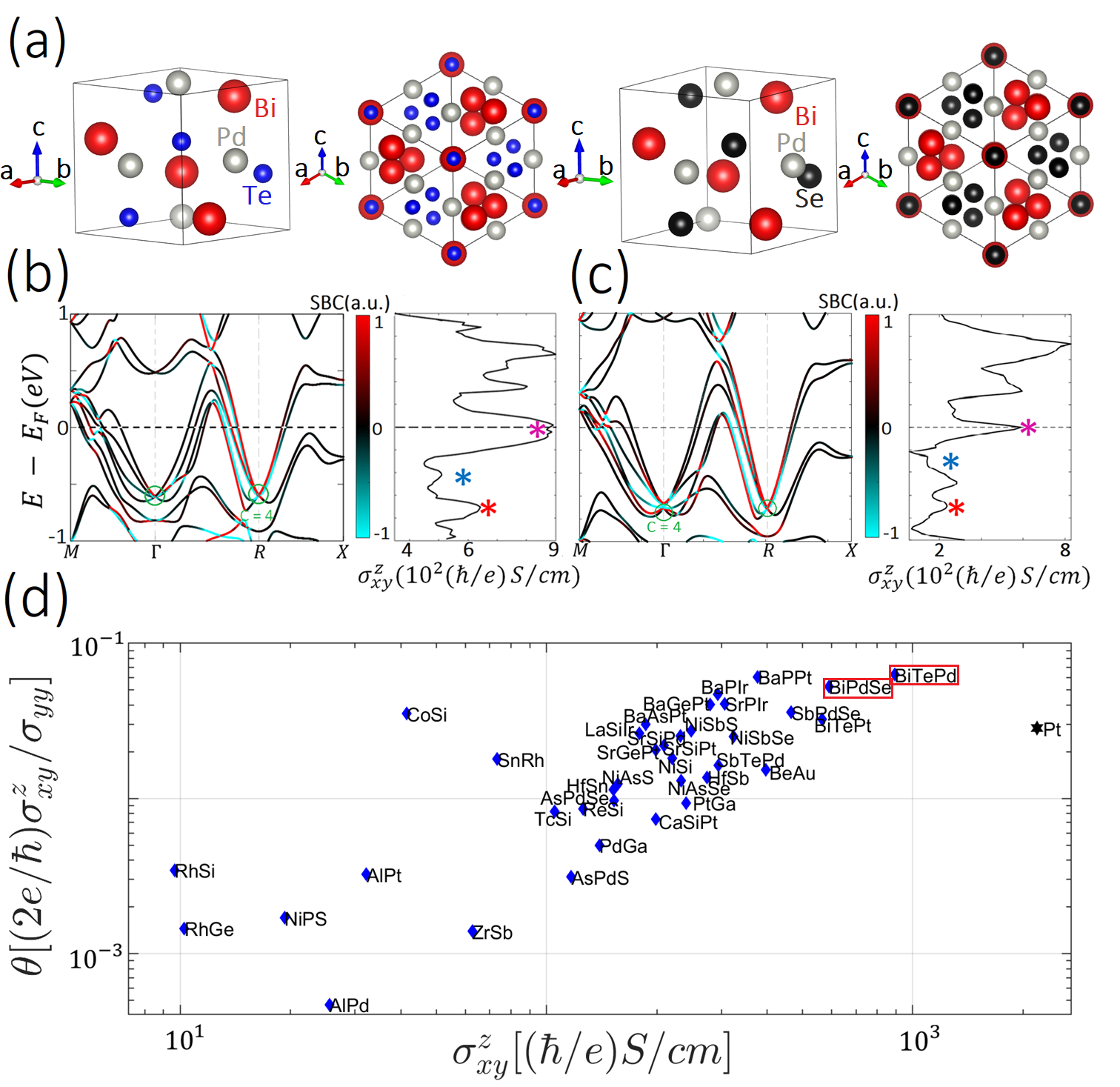}
 \caption{Crystal structures, band structures, SHCs, and SHAs of SG-198 CTSs. (a) Crystal structures of BiTePd and BiPdSe. (b) and (c), Band structure and SHC of BiTePd and BiPdSe, respectively. Colored asterisks correspond to same features in Fig. 3(b) center panel. (d) SHAs ($\theta$), and SHCs of 37 SG-198 materials at Fermi level with representatives highlighted in red boxes. The numerically determined SHA of Pt (black star) is included for comparison~\cite{SM}.}
\label{fig4}
\end{figure}

\paragraph{Conclusions and Perspectives.} We systematically investigated the intrinsic spin Hall effect (SHE) in topological semimetallic systems with high Chern numbers (CN), which suggests that SHE enhancement can occur under the energetic alignment of high CN Kramers-Weyl points (KWP). At $\mathcal{C} = 4$ KWP alignment, the spin Hall conductivity reveals a characteristic camel-back structure, whose stationary points can be traced to constructive contributions from intense spin Berry curvature hotspots near the high CN KWPs and SOC-induced nested bands emanating from these high degeneracy KWPs. 

It is important to note that the high CNs in the models studied here are mainly due to high band degeneracies. However, there are gapless systems with high CNs, which are due to higher-order band dispersions~\cite{Gilbert}. The SHE enhancement mechanism that we discovered in this work may differ in such systems. For a more complete understanding, systematic studies along these lines are essential.\\

\paragraph{Acknowledgments.}\begin{acknowledgments}C.O.A, W.J, D.S, S.L, J.P.W and T.L were partially supported by the SMART, one of seven centers of nCORE, a Semi- conductor Research Corporation program, sponsored by National Institute of Standards and Technology (NIST). D.S., J.P.W and T. L also acknowledge partial support from the DARPA ERI FRANC program under HR001117S0056-FP-042. S.L and T.L acknowledges partial support by the National Science Foundation through the University of Minnesota MRSEC under Award Number DMR-2011401. S.L. is also supported by Basic Science Research Program through the National Research Foundation of Korea (NRF) funded by the Ministry of Education (NRF-2021R1A6A3A14038837).
\end{acknowledgments}

%\bibliography{ISHC_CTSM} 
%\bibliography{ISHC_CTSM} 



\end{document}